\documentclass[aps,prb,twocolumn,groupedaddress,showpacs,showkeys,floatfix,superscriptaddress,10pt]{revtex4-1}

\usepackage{graphicx}
\usepackage{epsfig}

\usepackage{amssymb}
\usepackage{amsmath}
\usepackage{color}

\usepackage{sidecap}
\sidecaptionvpos{figure}{c}

\begin{document}

\title{Quantifying and controlling the magnetic dipole contribution \\ to 1.5 $\mu$m light emission in erbium-doped yttrium oxide}

\author{Dongfang Li}
\affiliation{School of Engineering, Brown University, Providence, RI 02912, USA}
\affiliation{Department of Physics, Brown University, Providence, RI 02912, USA}
\author{Mingming Jiang}
\affiliation{School of Engineering, Brown University, Providence, RI 02912, USA}
\affiliation{Department of Physics, Brown University, Providence, RI 02912, USA}
\author{S\'{e}bastien Cueff}
\altaffiliation{Present address: Institut des Nanotechnologies de Lyon, Ecole Centrale de Lyon, 69134 Ecully, France}
\affiliation{School of Engineering, Brown University, Providence, RI 02912, USA}
\author{Christopher M. Dodson}
\affiliation{School of Engineering, Brown University, Providence, RI 02912, USA}
\author{Sinan Karaveli}
\altaffiliation{Present address: Research Laboratory of Electronics, Massachusetts Institute of Technology, Cambridge, MA 02139, USA}
\affiliation{School of Engineering, Brown University, Providence, RI 02912, USA}
\author{Rashid Zia}
\email{Rashid\_Zia@brown.edu}
\affiliation{School of Engineering, Brown University, Providence, RI 02912, USA}
\affiliation{Department of Physics, Brown University, Providence, RI 02912, USA}

\date{\today}

\begin{abstract}
We experimentally quantify the contribution of magnetic dipole (MD) transitions to the near-infrared light emission from trivalent erbium-doped yttrium oxide (Er$^{3+}$:Y$_2$O$_3$). Using energy-momentum spectroscopy, we demonstrate that the $^4$I$_{13/2}{\to}^4$I$_{15/2}$ emission near 1.5 $\mu$m originates from nearly equal contributions of electric dipole (ED) and MD transitions that exhibit distinct emission spectra. We then show how these distinct spectra, together with the differing local density of optical states (LDOS) for ED and MD transitions, can be leveraged to control Er$^{3+}$ emission in structured environments. We demonstrate that far-field emission spectra can be tuned to resemble almost pure emission from either ED or MD transitions, and show that the observed spectral modifications can be accurately predicted from the measured ED and MD intrinsic emission rates. 
\end{abstract}
\pacs{32.50.+d, 32.70.-n, 42.50.Ct, 78.66.-w}

\maketitle
\vspace{-0.4cm}
\section{Introduction}
\vspace{-0.4cm}
Trivalent erbium ions are important light emitters for both fundamental and applied science. The near-infrared 1.5 $\mu$m $^4$I$_{13/2}{\to} ^4$I$_{15/2}$ transition in Er$^{3+}$ coincides with the low-loss window of silica fibers used in optical telecommunication. As a result, stimulated emission from Er$^{3+}$ is the basis for many fiber lasers and amplifiers.\cite{Digonnet} For decades, spontaneous emission from Er$^{3+}$ ions has been investigated for applications in integrated optics,\cite{Ennen,Zheng,Polman_JAP_1997,Iacona,Kenyon} including recent work exploring enhanced Er$^{3+}$ emission within optical nanostructures as the basis for chip-scale devices.~\cite{Barrios,Bao,Creatore,Yerci,Cueff_OpEx,Lawrence,Cueff_APL_2013} Spontaneous emission from Er$^{3+}$ also serves as the basis for the upconverting phosphors used in silicon-based near-infrared cameras,~\cite{CreaseyPatent} and recent work has shown how upconverting erbium-doped nanoparticles can serve as low-background, photo-stable light sources for microscopy and bio-imaging.\cite{Wu_PNAS,Chan_NanoLett,Chan_JPCB}    

In addition to its technological importance, Er$^{3+}$ light emission has been the subject of numerous scientific studies, including canonical experiments on the study of light-matter interactions. For example, erbium-doped materials were used in early demonstrations of controlled spontaneous emission through engineering the local density of optical states (LDOS).\cite{Vredenberg, Snoeks, Dood_PRA} Erbium emitters have also been used to investigate real and virtual cavity models for local field corrections.\cite{Zampredi_PRB} In recent theoretical works, the multipolar nature of the $^4$I$_{13/2}{\to} ^4$I$_{15/2}$ transition in Er$^{3+}$ has garnered new interest as a potential system with which to study and enhance magnetic light-matter interactions.\cite{Thommen_OL,Feng_OL,Rolly_PRB,Schmidt,Albella_JPCC,Hein_PRL} 

Although its emission is generally modeled as solely originating from forced electric dipole (ED) transitions,\cite{Vredenberg,Snoeks,Bao,Creatore,Dood_PRA} the  $^4$I$_{13/2}{\to}^4$I$_{15/2}$ transition in Er$^{3+}$ is directly allowed by magnetic dipole (MD) selection rules.\cite{Weber_1967} Theoretical studies suggest that Er$^{3+}$ emission near 1.5 $\mu$m originates from a combination of ED and MD transitions.\cite{Weber_1967,Weber_1968,Dammak,Dodson} Although an indirect lower bound can be inferred by comparing calculated MD emission rates with observed lifetimes,\cite{Weber_1968,Dodson} direct experimental quantification of the MD contribution to Er$^{3+}$ emission has been complicated by the degenerate nature of its ED and MD transitions. Such direct quantification is necessary if erbium-doped materials are to become an experimental test bed for studying magnetic light-matter interactions, or if the control over MD transitions, as recently demonstrated with trivalent europium and chromium,\cite{Karaveli_PRL, Karaveli_NanoLett,Karaveli_ACSNano} is to become an integral part of erbium technology. 

In this paper, we leverage energy-momentum spectroscopy\cite{Taminiau} to distinguish the spectrally overlapping ED and MD transitions in erbium-doped yttrium oxide (Er$^{3+}$:Y$_2$O$_3$), and thus provide direct measurements of the MD contribution in a model system for experimental investigation. Specifically, we demonstrate that the $^4$I$_{13/2}{\to}^4$I$_{15/2}$ emission near 1.5 $\mu$m originates from nearly equal contributions of ED and MD transitions that exhibit distinct emission spectra. By controlling the spacing between a gold mirror and an Er$^{3+}$:Y$_2$O$_3$ emitter layer, we modify the local optical environment and consequently alter the observed ratio of ED to MD emission in far-field spectra. Contrary to previous models for Er$^{3+}$ emission based on ED transitions, we show that these spectral modifications can only be modeled by considering the contributions of both ED and MD transitions. We also present analysis to help explain how this simple system can tune Er$^{3+}$ emission near the 1.5 $\mu$m telecommunication band to resemble either the ED or MD spectra, and then, briefly discuss the implications for nano-optical measurements and nanophotonic device design.
\vspace{-0.4cm}
\section{Quantifying ED and MD Emission}
\vspace{-0.4cm}
An erbium-doped emitter layer was deposited onto a quartz coverslip by sequential electron-beam evaporation of 94 nm Y$_2$O$_3$, 30 nm Er$^{3+}$-doped Y$_2$O$_3$ (3.5 at.~\% ), and 23 nm Y$_2$O$_3$.  The sample was subsequently annealed at 900 $^{\circ}\mathrm{C}$ under oxygen flow (0.5 lpm) for 45 minutes. Annealing, together with the top and bottom yttria buffer layers, helps to improve the emitter layer crystallinity, and the resulting emission closely resembles the published spectra for nanocrystalline Er$^{3+}$-doped yttria.\cite{Mao} (The emission spectra with and without the pure yttria buffer layers are shown in Supplemental Figure S1.\cite{Supp_Mat})

To quantify the ED and MD contributions to light emission, we use energy-momentum spectroscopy.\cite{Taminiau} This technique measures the angular distribution of light emission as a function of wavelength, and distinguishes ED and MD emission by their distinct radiation patterns. Here, the energy-momentum spectra were obtained from the Er$^{3+}$:Y$_2$O$_3$ layer under continuous excitation by a 532 nm laser (Coherent Verdi) with a setup similar to that reported in Ref.~\onlinecite{Karaveli_arXiv}. An inverted microscope (Nikon TE2000) with an oil-immersion objective (100x, 1.3 NA) was used to both pump the emitters and collect their fluorescence. The back focal plane of the objective was projected through a linear polarizer onto the entrance slit of an imaging spectrograph (Princeton Instruments IsoPlane) using a 100 mm focal length Bertrand lens. This polarized Fourier image was dispersed by the spectrograph, and the resulting energy- and momentum-resolved spectra were recorded by a 2D InGaAs focal plane array (Princeton Instruments NIRvana). 

\begin{figure}[t]
  \includegraphics[width=3.5in]{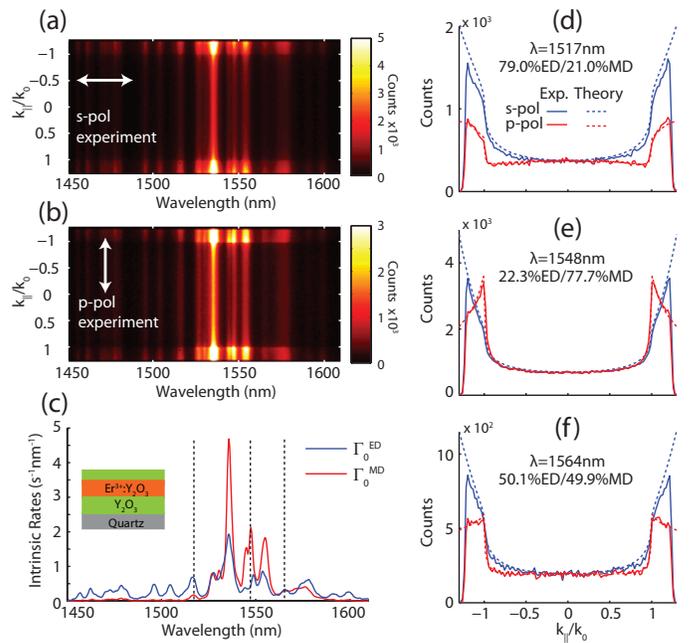}
  \caption{(Color online) Quantifying the near-infrared ED and MD transitions in Er$^{3+}$:Y$_2$O$_3$ by energy-momentum spectroscopy. (a, b) Experimental energy-momentum spectra obtained for s- and p-polarization, respectively. (c) Spectrally resolved emission rates, ${\Gamma }_0^{ED}$ (blue) and ${\Gamma }_0^{MD}$ (red), obtained from fitting the experimental momentum cross sections at each wavelength. Inset shows a schematic of the measured sample. (d-f) Comparison of experimental data and theoretical fits for three peak wavelengths (1517, 1548, and 1564 nm) as highlighted by dashed lines in (c).}
  \label{fig1}
\end{figure}

Figures~\ref{fig1}(a) and~\ref{fig1}(b) show the measured energy-momentum spectra for s- and p-polarization, respectively. Figure ~\ref{fig1}(c) shows the spectrally resolved ED and MD intrinsic emission rates extracted from theoretical fits to the p-polarized data. The relative rates were first obtained using the methods previously described in Ref.~\onlinecite{Taminiau} and then converted to absolute units by normalizing the integrated MD rate over the whole $^4$I$_{13/2}{\to} ^4$I$_{15/2}$ transition to the value obtained from free-ion calculations in Ref.~\onlinecite{Dodson} (${\int{\Gamma_0^{MD}}d\lambda}=1.74^3~\times 10.17~s^{-1}=53.6~s^{-1}$).\cite{Sum_Rule} 

The measured emission rates in Fig.~\ref{fig1}(c) reveal distinct ED and MD spectra. Whereas the MD emission is concentrated between 1525 and 1565 nm, the ED emission is more evenly distributed between 1450 and 1610 nm. Despite the strongly mixed nature of the overall emission, individual emission peaks tend to be dominated by either ED or MD transitions. To highlight this effect and also to demonstrate the accuracy of the theoretical fits, Figs.~\ref{fig1}(d)-\ref{fig1}(f) show the experimental and theoretical momentum-resolved cross sections at three peak wavelengths. At 1517 nm, ED transitions account for 79\% of the intrinsic emission, whereas 1548 nm emission is dominated by over 77\% MD transitions. There are only a few peaks where emission from ED and MD transitions are comparable, such as 1564 nm. Integrating the intrinsic emission rates in Fig.~\ref{fig1}(c) over the full spectral range from 1450 to 1610 nm, we find that MD transitions account for 49.8\% of the total intrinsic $^4$I$_{13/2}{\to} ^4$I$_{15/2}$ emission in Er$^{3+}$:Y$_2$O$_3$. These results are consistent with the 47\% lower bound on the MD contribution, which can be inferred from the observed lifetime and calculated MD emission rates reported in Ref.~\onlinecite{Weber_1968}. (The directly measured 49.8\% value is also within the 33\% to 59\% MD contribution range, which can be inferred from the Judd-Ofelt analysis in Ref.~\onlinecite{Weber_1968}.) These experimental results thus reaffirm theoretical predictions that Er$^{3+}$:Y$_2$O$_3$ is a strongly mixed ED and MD emitter, and by characterizing the distinct ED and MD contributions, may open new opportunities for the design of Er$^{3+}$-doped devices. 

\vspace{-0.4cm}
\section{Modifying ED and MD Contributions}
\vspace{-0.4cm}
The strong and distinct contribution of MD transitions can provide a method to tune the near-infrared emission of Er$^{3+}$-doped thin films. As demonstrated previously with the spectrally distinct $^5$D$_0{\to}^7$F$_1$ MD and $^5$D$_0{\to}^7$F$_2$ ED transitions in Eu$^{3+}$, changes to the local optical environment can tune emission spectra by selectively enhancing either ED or MD emission.\cite{Karaveli_PRL} Here, we use a similar experimental configuration to show how changes to the LDOS in close proximity to a gold mirror can be used to modify the relative ED and MD contributions to the $^4$I$_{13/2}{\to} ^4$I$_{15/2}$ emission from Er$^{3+}$:Y$_2$O$_3$. 

\begin{SCfigure*}
  \centering
  \includegraphics[width=12.9cm]{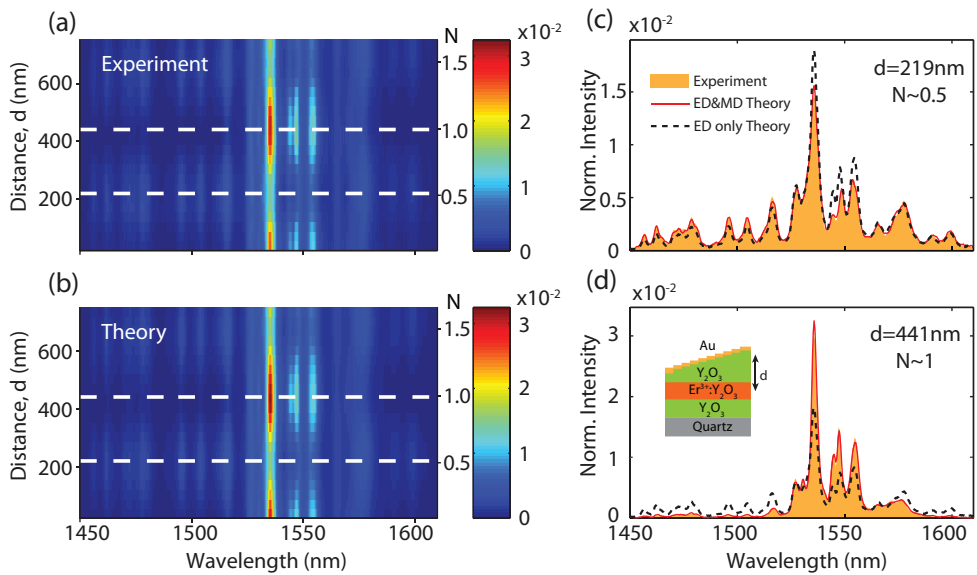}
  \caption{(Color online) LDOS-based tuning of Er$^{3+}$:Y$_2$O$_3$ emission near a gold mirror. (a) Normalized experimental spectra measured for different emitter-mirror separation distances, $d$; each spectra is normalized to its total integrated intensity.\cite{Supp_Mat} (b) Normalized theoretical spectra predicted by the three-layer LDOS model using ED and MD intrinsic rates from Fig.~1(c). (c,~d) Comparison of experimental and theoretical spectra for $d=219$ nm and $d=441$ nm as indicated by dashed white lines in (a,~b). Inset in panel (d) shows a schematic of the sample with different thickness spacer layers between the emitter layer and gold mirror.}
  \label{fig2}
\end{SCfigure*}

To this end, we deposited a staircase-like structure comprising twenty-two different thickness spacer layers of Y$_2$O$_3$ (in $\sim$33 nm intervals from 0 to 697 nm) by masked electron-beam evaporation on top of the aforementioned emitter sample measured in Fig.~\ref{fig1}. (The resulting spacer layer thicknesses were monitored during deposition using a quartz oscillator and subsequently confirmed by ellipsometry on a silicon reference sample.) On top of the staircase, we then deposited a 5 nm Ti adhesion layer and 200 nm gold film by electron-beam evaporation. 

Figure~\ref{fig2}(a) shows how the experimental spectra change with the emitter-mirror separation distance, $d$, as measured from the center of the Er$^{3+}$:Y$_2$O$_3$ emitter layer. This 2D plot displays the far-field emission spectra measured from all twenty-two steps through the quartz substrate using a 60x, 0.85 NA objective under continuous 532 nm laser excitation. As $d$ increases, the normalized emission spectra show periodic oscillations between regions where it resembles either the MD or ED intrinsic emission rates shown in Fig.~1(c). The experimental spectra vary from being concentrated in several pronounced peaks near 1545 nm and being spread out more evenly over a dozen peaks between 1450 and 1610 nm. 

The observed spectral changes can be accurately predicted by modeling how the different LDOS for ED and MD transitions changes with emitter-mirror separation distance.\cite{Karaveli_PRL,Karaveli_NanoLett} Figure~\ref{fig2}(b) shows the normalized theoretical emission spectra obtained using the intrinsic emission rates shown in Fig.~1(c). To model the LDOS variations, we use the three-layer interference model from Chance, Prock, and Silbey,\cite{Chance} but unlike prior work with spectrally distinct ED and MD transitions,\cite{Karaveli_PRL,Karaveli_NanoLett} we must explicitly consider the overlapping ED and MD contributions in Er$^{3+}$. Therefore, we model the emission rate at each wavelength $\lambda$ and distance $d$ as ${\Gamma }(\lambda,d)={\Gamma }_0^{ED}(\lambda)\tilde{\rho}^{ED}(\lambda,d)+{\Gamma }_0^{MD}(\lambda)\tilde{\rho}^{MD}(\lambda,d)$. Here, ${\Gamma }_0^{ED}$ and ${\Gamma }_0^{MD}$ are the spectrally resolved ED and MD intrinsic emission rates obtained by energy-momentum spectroscopy, which must be scaled by the appropriate ED and MD LDOS at the emitter location, $\tilde{\rho}^{ED}$ and $\tilde{\rho}^{MD}$.\cite{Supp_Mat} 

Figures~\ref{fig2}(c) and~\ref{fig2}(d) compare the experimental data and theoretical predictions for the two different emitter-mirror distances, 219 and 441 nm, denoted by the dashed white lines in Figs. 2(a) and 2(b). The theoretically predicted spectra (solid red lines), which take into account the mixed ED and MD nature of emission, show excellent agreement with the normalized experimental spectra (shaded orange regions). To emphasize that the spectral changes could not be predicted without considering the MD contribution, we also plot theoretical spectra assuming that all emission originates from ED transitions (dashed black lines). As the emission bandwidth ($\sim$150 nm) is small compared to the center emission wavelength ($\sim$ 1.5 $\mu$m), the ED only model predicts almost no change in spectral shape (see Supplemental Figure S2).\cite{Supp_Mat} Therefore, one must consider both ED and MD transitions to accurately predict the observed spectral variation. 

To quantify the extent of the spectral variation, we can calculate what fraction of the predicted spectra originates from ED and MD transitions. For example, Figs. 3(a) and 3(b) show the distinct ED and MD contributions to the far-field emission spectra for the $d$ = 219 and 441 nm cases, respectively. Note that for the 219 nm case, emission appears to be dominated by ED transitions; the fractional MD contribution to the integrated emission spectra, as calculated by the branching ratio $a_{MD}={\int{\Gamma^{MD}}d\lambda}/{\int{\left( {{\Gamma }^{ED}}+{{\Gamma }^{MD}} \right)d\lambda }}$, is only 7.8\%. In contrast, the emission for the 441 nm case is dominated by MD transitions: $a_{MD}=$ 91.3\%. Thus, for this highly mixed ED and MD emitter system, we can selectively tune over 90\% of the total emission to either ED or MD transitions using a mirror and proper spacer layer.

\begin{SCfigure*}
\centering
  \includegraphics[width=4.8in]{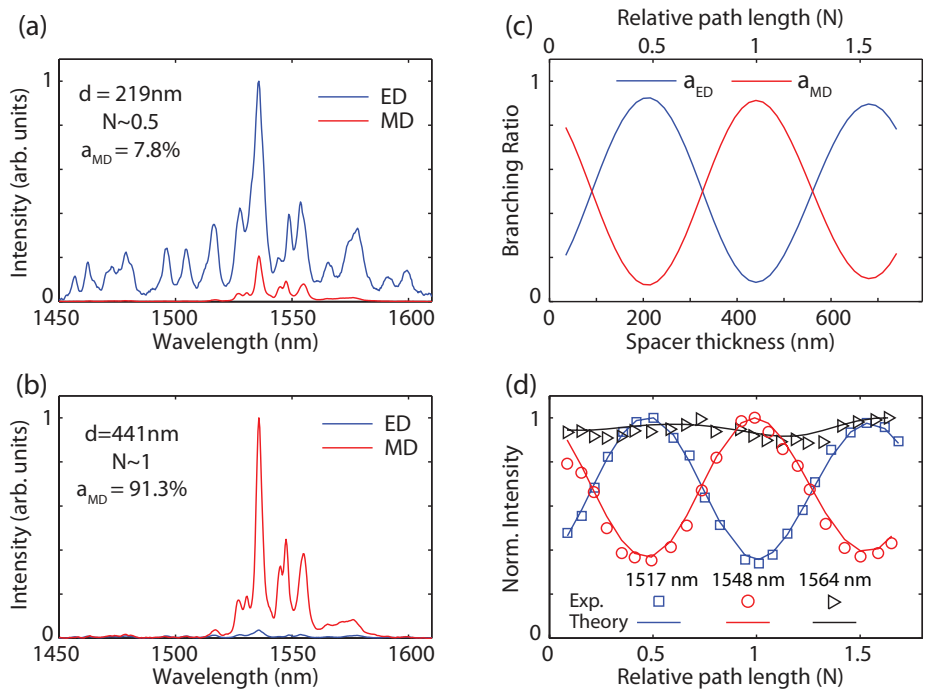}
\caption{(Color online) Analysis of ED and MD contributions. (a,~b) Relative ED and MD contributions to the emission spectra as predicted by theoretical calculations for the $d$ = 219 and 441 nm cases. (c) Fractional contribution of the ED and MD transitions integrated over the full spectral range from 1450 to 1610 nm. (d) Comparison of the experimentally observed and theoretically predicted normalized intensity variations at three representative wavelengths: 1517 nm (blue), 1548 nm (red), and 1564 nm (black) as a function of relative path length, N. The intensity traces in panel (d) present vertical cross-sections through Figs.~2(a) and 2(b), in which each cross-section has been normalized to its maximum value to facilitate side-by-side comparison.}
  \label{fig3}
\end{SCfigure*}

The strong variations in the ED and MD contributions are a direct result of the near-equal integrated ED and MD intrinsic emission rates. Since the intrinsic ED and MD contributions are balanced, changes to the LDOS can easily promote either pathway to emission. To demonstrate this effect, we plot in Fig.~3(c) the fractional ED and MD contributions, $a_{ED}$ and $a_{MD}$, as a function of emitter-mirror distance, $d$. Both of these contributions periodically oscillate around a mean value near 50\%. To intuitively understand the variation with distance, we define a round-trip relative path length in terms of wavelength units: N $\equiv 2 n_r d/\lambda$, where $n_r$=1.74 is the refractive index of the spacer layer. The top label across Fig.~3(c) shows the relative path length N in terms of the central peak wavelength $\lambda = 1536$ nm. It is clear that the MD contribution is enhanced for integer values of N, whereas the ED contribution is enhanced for half-integer values. This is consistent with an intuitive model of ED and MD enhancement in terms of interference near the mirror of the radiated electric and magnetic fields, respectively. For a perfect metal mirror, the phase of the reflected electric field is shifted by $\pi$. Therefore, for an ED transition, one would expect destructive interference to suppress emission at integer values of N, while constructive interference would enhance ED emission at half-integer values. In contrast, the reflected magnetic field from a perfect mirror acquires no phase shift, and thus, the opposite should occur for MD transitions. MD emission should be enhanced at integer values of N and suppressed at half-integer values. 

This simple path length model can also be used to understand the relative intensity changes at each wavelength. In Fig.~3(d), we plot the experimentally observed and theoretically predicted intensity variations as a function of N for the three peak wavelengths considered earlier in this paper. At wavelengths where either ED or MD transitions dominate, such as 1517 and 1548 nm, the intensity follows the trend of the ED or MD curves in Fig.~3(c). However, when the spectrally resolved intrinsic ED and MD emission rates are nearly equal, as is the case at 1564 nm, there are no significant variations in the intensity contribution with path length. (Minor variations observed in the normalized intensity at 1564 nm result from dispersion effects and small asymmetries in ED and MD rates.) In this way, the normalized intensity variation at each wavelength can be approximated by simply scaling the ED and MD curves in Fig.~3(c) by the appropriate intrinsic ED and MD rates from Fig.~1.
\vspace{-0.4cm}
\section{Conclusion}
\vspace{-0.4cm}
In conclusion, we have experimentally quantified the ED and MD contributions to the $^4$I$_{13/2}{\to} ^4$I$_{15/2}$ transition in Er$^{3+}$:Y$_2$O$_3$. Energy-momentum spectroscopy measurements demonstrated that MD transitions account for $\sim$50\% of the intrinsic emission around 1.5 $\mu$m in this system, and more importantly, provided a direct means to distinguish the spectrally resolved ED and MD intrinsic emission rates. These measurements revealed distinct spectra for ED and MD emission, which were then leveraged to tune the far-field emission spectra in close proximity to a gold mirror. We showed that the observed spectral modifications could be accurately predicted from the measured ED and MD intrinsic rates using analytical calculations of the LDOS in a multilayer system. Furthermore, we showed that the observed spectral variations are consistent with a simple model based on relative path length changes with emitter location. 

The change of observed spectra with emitter location suggests a range of device designs that could be used to modulate Er$^{3+}$ emission in the 1.5 $\mu$m telecommunication band. As recent experiments with Eu$^{3+}$ have shown,\cite{Karaveli_NanoLett} dynamic control of the LDOS can modify emission spectra at sub-lifetime speeds. Thus active structures, which rapidly vary the  emitter position or relative path length, could help pave the way for new high-speed directly modulated erbium light-emitting devices. 

From a broader scientific perspective, the results presented here help define a new system with which to study magnetic light-matter interactions. In particular, the spectrally resolved emission rates for Er$^{3+}$:Y$_2$O$_3$ could help guide new experiments to test recent theoretical predictions for the modified ED and MD emission rates near optical nanostructures, including metallic antennas,\cite{Taminiau_NanoLett,Feng_OL} dielectric particles,\cite{Rolly_PRB,Schmidt,Albella_JPCC} and split-ring resonators.\cite{Hein_PRL} As compared to shorter wavelength MD transitions in Cr$^{3+}$ and Eu$^{3+}$, the near-infrared MD transitions in Er$^{3+}$ provide a much more practical system for which to design and fabricate resonant nanostructures. The methods presented here could also be applied to study the MD transitions in other promising host materials and ions.\cite{Dodson}

\vspace{-0.4cm}
\begin{acknowledgments}
\vspace{-0.4cm}
The authors thank J. A. Kurvits for helpful discussions. Financial support for this work was provided by the Air Force Office of Scientific Research (PECASE award FA9550-10-1-0026) and the National Science Foundation (CAREER award EECS-0846466).
\end{acknowledgments}
\vspace{-0.4cm}
\bibliography{ErReference}

\end{document}